\documentstyle[twocolumn,aps,psfig]{revtex}

\begin{document}
\draft
\flushbottom
\twocolumn[
\hsize\textwidth\columnwidth\hsize\csname @twocolumnfalse\endcsname

\title{ Linear and nonlinear optics of surface plasmon toy-models of black holes and wormholes.}
\author{Igor I. Smolyaninov}
\address{ Department of Electrical and Computer Engineering \\
University of Maryland, College Park,\\
MD 20742}
\date{\today}
\maketitle
\tightenlines
\widetext
\advance\leftskip by 57pt
\advance\rightskip by 57pt

\begin{abstract}
Experimental and theoretical studies of linear and nonlinear optics of surface plasmon toy wormholes and black holes have been performed. These models are based on dielectric microdroplets on the metal surfaces and on nanoholes drilled in thin metal films. Toy surface plasmon black holes and wormholes are shown to exhibit strongly enhanced nonlinear optical behavior in the frequency range near the surface plasmon resonance of a metal-liquid interface. Various possibilities to emulate such nontrivial gravitation theory effects as Hawking radiation and Cauchy horizons are discussed. 
\end{abstract}

]
\narrowtext

\tightenlines

I. INTRODUCTION.

The realization that solid-state toy models may help in an understanding of electromagnetic phenomena in curved space-time has led to considerable recent effort in developing toy models of electromagnetic \cite{1,2} and sonic \cite{3,4,5} black holes. In the case of media electrodynamics this is possible because of an analogy between the propagation of light in matter and in curved space-times: it is well known that Maxwell equations in a general curved space-time background $g_{ik}(x,t)$ are equivalent to the phenomenological Maxwell equations in the presence of a matter background with nontrivial electric and magnetic permeability tensors $\epsilon _{ij}(x,t)$ and $\mu _{ij}(x,t)$ \cite{1,6}. In this analogy, the event horizon corresponds to a surface of singular electric and magnetic permeabilities, so that the speed of light goes to zero, and light is "frozen" near such a surface. In the absence of established quantum gravitation theory the toy models are helpful in understanding electromagnetic phenomena in curved space-times, such as Hawking radiation \cite{7} and the Unruh effect \cite{8}. Unfortunately, up to now all the suggested electromagnetic black hole toy-models were very difficult to realize and study experimentally, so virtually no experimental work has been done in this field. 

In this paper we propose and study novel electromagnetic toy models of wormholes and black holes, which are based on surface plasmons and are very easy to make and explore in the experiment. We are going to consider a close analogy between the optics of surface plasmons propagating along metal interfaces, which are curved and/or covered with dielectric films, and the field theory in a curved space-time background. Surface plasmons are collective excitations of the conductive electrons and the electromagnetic field \cite{9}. They exist in "curved three-dimensional space-times" defined by the shape of the metal-dielectric interface. Since in many experimental geometries surface plasmons are weakly coupled to the outside world (to free-space photons, phonons, single-electron excitations, etc.) it is reasonable to treat the physics of surface plasmons separately from the rest of the surface and bulk excitations, so that a field-theory of surface plasmons in a curved space-time background may be considered. For example, a nanohole in a thin metal membrane may be treated as a "wormhole" connecting two "flat" surface plasmon worlds located on the opposite interfaces of the membrane \cite{10}. On the other hand, near the plasmon resonance (which is defined by the condition that $\epsilon _m(\omega )=-\epsilon _d$ at the metal-dielectric interface, where $\epsilon _m(\omega )$ and $\epsilon _d$ are the dielectric constants of metal and dielectric, respectively \cite{9}) the surface plasmon velocity vanishes, so that the surface plasmon "stops" on the metal surface, and the surface charge and the normal component of the electric field diverge. For a given frequency of light, the spatial boundary of the plasmon resonance ("the event horizon" of our toy model) may be defined at will using the geometry $\epsilon _d(x,y)$ of the absorbed layer of dielectric on the metal surface. Thus, the plasmon resonance becomes a natural candidate to emulate the event horizon of a black hole. As a result, toy two-dimensional surface plasmon black holes can be easily produced and studied. In what follows we are going to show that consideration of such surface plasmon toy wormholes and black holes may be useful both for the fields of nanooptics and for gravitation theory. 

Let us consider in detail the dispersion law of a surface plasmon (SP), which propagates along the metal-dielectric interface. The SP field decays exponentially both inside the metal and the dielectric. Inside the dielectric the decay exponent is roughly equal to the SP wave vector. As a first step let us assume that both metal and dielectric completely fill the respective $z<0$ and $z>0$ half-spaces. In such a case the dispersion law can be written as \cite{9} 

\begin{equation}  
k^2=\frac{\omega ^2}{c^2}\frac{\epsilon _d\epsilon _m(\omega )}{\epsilon _d+\epsilon _m(\omega)} ,
\end{equation}

where we will assume that $\epsilon _m=1-\omega _p^2/\omega ^2$ according to the Drude model, and $\omega _p$ is the plasma frequency of the metal. This dispersion law is shown in Fig.1(b) for the cases of metal-vacuum and metal-dielectric interfaces. It starts as a "light line" in the respective dielectric at low frequencies and approaches asymptotically $\omega =\omega _p/(1+\epsilon _d)^{1/2}$ at very large wave vectors. The latter frequency corresponds to the so-called surface plasmon resonance. Under the surface plasmon resonance conditions both phase and group velocity of the SPs is zero, and the surface charge and the normal component of the electric field diverge. Since at every wavevector the SP dispersion law is located to the right of the "light line", the SPs of the plane metal-dielectric interface are decoupled from the free-space photons due to the momentum conservation law.   

If a droplet of dielectric (Fig.1(a)) is placed on the metal surface, the SP dispersion law will be a function of the local thickness of the droplet. Deep inside the droplet far from its edges the SP dispersion law will look similar to the case of a metal-dielectric interface, whereas near the edges (where the dielectric is thinner) it will approach the SP dispersion law for the metal-vacuum interface. As a result, for every frequency between $\omega _p/(1+\epsilon _d)^{1/2}$ and $\omega _p/2^{1/2}$ there will be a closed linear boundary inside the droplet for which the surface plasmon resonance conditions are satisfied (Fig.1(b,c)). Let us show qualitatively that such a droplet of dielectric on the metal interface behaves as a "surface plasmon black hole" in the frequency range between $\omega _p/(1+\epsilon _d)^{1/2}$ and $\omega _p/2^{1/2}$, and that the described boundary of the surface plasmon resonance behaves as an "event horizon" of such a black hole. Let us consider a SP within this frequency range, which is trapped near its respective "event horizon", and which is trying to leave a large droplet of dielectric (Fig.1(c)). The fact that the droplet is large means that ray optics may be used. Since the component of the SP momentum parallel to the droplet boundary has to be conserved, such a SP will be totally internally reflected by the surface plasmon resonance boundary back inside the droplet at any angle of incidence. This is a simple consequence of the fact that near the "event horizon" the effective refractive index of the droplet for surface plasmons is infinite (according to eq.(1), both phase and group velocity of surface plasmons is zero at surface plasmon resonance). On the other hand, all the incoming plasmons in this frequency range will be "sucked" into the droplet. Thus, the droplet behaves as a black hole for surface plasmons, and the line near the droplet boundary where the surface plasmon resonance conditions are satisfied plays the role of the event horizon for surface plasmons. 

II. EFFECTIVE METRICS OF SURFACE PLASMON BLACK HOLES. 

Quantitatively the effective metric experienced by surface plasmons near the droplet boundary may be written by introducing an effective slowly varying local SP phase velocity $c^\star (x,y) =\omega /k(x,y)$, so that the wave equation for surface plasmons may be written as 

\begin{equation}
(\frac{\partial ^2}{c^{\star 2}\partial t^2}-\frac{\partial ^2}{\partial x^2}-\frac{\partial ^2}{\partial y^2})A=0,
\end{equation}

which corresponds to an effective metric

\begin{equation}
ds^2=c^{\star 2}dt^2-dx^2-dy^2
\end{equation}

The behavior of $c^\star (x,y)$ is defined by the shape and thickness of the droplet near its edge (if necessary, the droplet may be replaced by a similar shaped layer of solid dielectric), and by the thickness of the metal film. 
In order to simplify situation, let us consider a thin metal membrane with two "linear" droplets positioned symmetrically on both sides of the membrane (Fig.2(a)). The droplets thicknesses at $x=0$ correspond to the surface plasmon resonance at the illumination frequency. The droplets taper off adiabatically in positive and negative $x$-directions on both sides of the membrane. In the symmetric membrane geometry the surface plasmon spectrum consists of two branches $\omega _-$ and $\omega _+$, which exhibit positive and negative dispersion, respectively, near the surface plasmon resonance \cite{11}. Fig.2(b) shows the dispersion curves of both branches for the cases of metal-vacuum interface far from the droplets, and for the locations near $x=0$. The droplets represent an effective black hole for $\omega _-$ modes and an effective white hole for $\omega _+$ modes, similar to \cite{1}. 

The event horizon corresponds to $c^\star =0$ at $x=0$. The two regions $x>0$ and $x<0$ correspond to the two sides of a black hole, which are connected by an Einstein-Rosen bridge at $x=0$. The behavior of $c^\star (x)$ near $x=0$ may be defined at will by choosing the corresponding geometry of the droplet edge. In order to adhere to the metric considered in \cite{1} let us assume that $c^\star =\alpha xc$ in the vicinity of $x=0$. However, we should remember that this linear behavior will be cut off somewhere near the effective horizon due to such effects as Landau damping \cite{12}, losses in the metal and the dielectric, etc. The resulting effective metric now looks like

\begin{equation}
ds^2=\alpha ^2x^2c^2dt^2-dx^2-dy^2
\end{equation}

Thus, this particular choice of the shape of the droplet edge gives rise to an effective Rindler geometry. Similar metric would describe the space-time geometry near the event horizon of a black hole with mass $M_{BH}=c^2/8\gamma \alpha $, where $\gamma $ is the gravitation constant \cite{1}. 
Choosing a different behavior of $c^\star (x)$ near $x=0$ leads to a different effective metric. For example, the choice of $c^\star =\beta x^2c$ would emulate the metric of an extremal electrically charged Reissner-Nordstrom black hole \cite{6} with mass $M=Q/\gamma ^{1/2}$ near its horizon, where $Q$ is the electric charge of the black hole. These various choices of the effective metric may be made possible by choosing the appropriate dielectric constant $\epsilon _d(x,y)$ and/or thickness $d(x,y)$ distribution of the absorbed layer of dielectric over the metal surface. The effects of these factors on the phase velocity of surface plasmons are complimentary to each other, which may be seen from the dispersion law of surface plasmons in a three-layer metal-dielectric-vacuum geometry described in numerous publications (see for example \cite{9}). In the limit of small plasmon phase velocities $c^{\star }/c<<1$ the phase velocity can be found by solving the following nonlinear equation:

\begin{equation}
\frac{c^{\star 2}}{c^2}=\eta = (\frac{1}{\epsilon _d}+\frac{1}{\epsilon _m})+\frac{(\epsilon _d-1)}{(\epsilon _d+1)}(\frac{1}{\epsilon _d}-\frac{1}{\epsilon _m})exp(-\frac{2\omega d}{c\eta ^{1/2}}) ,
\end{equation}

so that one can vary either $\epsilon _d$ or $d$ in order to achieve the same desired phase velocity at a given frequency $\omega $ and at a given location on the metal surface.

Finally, our model can be extended in order to emulate the space-time metric of a rotating black hole (Kerr metric \cite{13}). Physical realization of this model involves a droplet of an optically active liquid on the metal surface which supports propagation of surface plasmons. As a result of this analogy, the droplets of such liquids on the metal surfaces may exhibit giant optical activity in the frequency range near the surface plasmon resonance of a metal-liquid interface.  

The relationship between the $\vec{D}$ and $\vec{E}$ fields in a gravitational field with the metric $g_{ik}$ can be written in the form \cite{13}:

\begin{equation}  
\vec{D}=\frac{\vec{E}}{h^{1/2}}+\vec{H}\times \vec{g} ,
\end{equation}

where $h=g_{00}$ and $g_{\alpha }=-g_{0\alpha }/g_{00}$. For an electromagnetic wave this relationship is very similar to the relationship between the $\vec{D}$ and $\vec{E}$ fields in the optically active medium, if we identify $1/h^{1/2}$ as $\epsilon $ of the medium, and recall how optical activity is introduced in media electrodynamics.  

There are different ways of introducing optical activity (gyration) tensor in the macroscopic Maxwell equations. It can be introduced in a symmetric form, which is sometimes called Condon relations \cite{14}:

\begin{equation}  
\vec{D}=\epsilon \vec{E}+\gamma\frac{\partial \vec{B}}{\partial t}
\end{equation}

\begin{equation}  
\vec{H}=\mu ^{-1}\vec{B}+\gamma\frac{\partial \vec{E}}{\partial t}
\end{equation}

Or it can be introduced only in an equation for $\vec{D}$ (see \cite{15,16}). In our consideration we will follow Landau and Lifshitz \cite{15}, and for simplicity use only the following equation valid in isotropic or cubic-symmetry materials:

\begin{equation}  
\vec{D}=\epsilon \vec{E}+i\vec{E}\times \vec{g} ,
\end{equation}

where $\vec{g}$ is called the gyration vector. If the medium exhibits magneto-
optical effect and does not exhibit natural optical activity $\vec{g}$ is 
proportional to the magnetic field $\vec{H}$:

\begin{equation}  
\vec{g}=f\vec{H} ,
\end{equation}

where the constant $f$ may be either positive or negative. For metals in the 
Drude model at $\omega >>eH/mc$

\begin{equation}  
f(\omega )= -\frac{4\pi Ne^3}{cm^2\omega ^3}=-\frac{e\omega _p^2}{mc\omega ^3} ,
\end{equation}

where $\omega _p$ is the plasma frequency and $m$ is the electron mass \cite{13}. In any case, the local difference in refractive index for the left $n_-$ and right $n_+$ circular polarizations of light is proportional to the local value of the gyration vector $\vec{g}$.

Thus, from the relationships between the $\vec{D}$ and $\vec{E}$ fields in a gravitational field and in media electrodynamics we conclude that space-time behaves as a chiral optical medium if $\vec{g}\neq 0$ (we must take into account different conventions for the use of imaginary numbers implemented in \cite{13} and \cite{15}, and hence equations (6) and (9)).

A space-time region with $\vec{g}\neq 0$ can be described locally as a rotating coordinate frame with an angular velocity \cite{13}

\begin{equation}  
\vec{\Omega }=\frac{ch^{1/2}}{2}\vec{\nabla }\times \vec{g} 
\end{equation}
 
Similar vector $\vec{\Omega }$ field can be defined for any chiral medium regardless of the nature of its optical activity (natural or magnetic field induced). Thus, electrodynamics of such medium may be understood as if there is a distribution of local angular rotation $\vec{\Omega }$ field inside the medium.

Now it is clear that in order to emulate a rotating black hole we must use a droplet of liquid which exhibits optical activity. Let us consider a medium with $\vec{g}=\vec{g}(r)$ directed along the $\phi $-coordinate (Fig.3(b)). Such a distribution may be produced, for example, in a twisted layer of liquid crystal or in a liquid droplet exhibiting magneto-optical effect if a droplet is formed around a small cylindrical wire and a current is passed through the wire. Such a geometry will result in different refractive indices $n_+$ and $n_-$ of the liquid as seen by the surface plasmons, which propagate in the clockwise (right) and counterclockwise (left) directions near the droplet boundary. As a result, these plasmons will have slightly different dispersion laws (Fig.3(a)). This difference produces two main effects: the position of the effective event horizon for the left and right plasmons will be different inside the liquid droplet (Fig.3(b)); and there will be a narrow frequency range (located near the surface plasmon resonance of the metal-liquid interface) where the event horizon will exist for only one (left or right) kind of surface plasmons (Fig.3(a)). 

The first effect closely resembles the properties of the Kerr metric, which characterizes a rotating black hole. The Kerr metric has two characteristic surfaces around a rotating black hole. There exists a spherical event horizon similar to the case of a non-rotating black hole, and there is a so called ergosphere \cite{13} just outside the black hole event horizon, which corresponds to the area of space where every particle must rotate in the direction of the black hole rotation. There are no particle states, which rotate in the opposite direction inside the ergoshere. Similar to the case of a real rotating black hole, our toy surface plasmon black hole has a toy-ergosphere: between the two event horizons shown in Fig.3(b) there are no surface plasmon states which rotate in the direction opposite to the direction of the effective $\vec{\Omega }$ field of the toy black hole. This area of space is quite unusual from the point of view of surface plasmons. We may say that even the surface plasmon vacuum rotates in this area, similar to the description of the electromagnetic vacuum inside the band gap of a photonic crystal as being "colored" vacuum. Further study of the linear and nonlinear optical properties of such rotating vacuum may be quite interesting from the fundamental optics point of view.

The second effect may be quite important from the point of view of practical applications. In the frequency range where the effective event horizon exists only for one (left or right) kind of plasmons, the array of droplets will behave as a medium which exhibits giant planar optical activity. Such media are being actively developed at the moment \cite{17}.  

III. EXPERIMENTAL OBSERVATIONS. 

The toy black holes described above are extremely easy to make and observe. In our experiments a small droplet of glycerin was placed on the gold film surface and further smeared over the surface using lens paper, so that a large number of glycerin microdroplets were formed on the surface (Fig.4(a)). These microdroplets were illuminated with white light through the glass prism (Fig.1(a)) in the so-called Kretschman geometry \cite{9}. The Kretschman geometry allows for efficient SP excitation on the gold-vacuum interface due to phase matching between the SPs and photons in the glass prism. As a result, SPs were launched into the gold film area around the droplet. Photograph taken under a microscope of one of such microdroplets is shown in Fig.4(b). The white rim of light near the edge of the droplet is clearly seen. It corresponds to the effective SP event horizon described above. Near this toy event horizon SPs are stopped or reflected back inside the droplet. In addition, a small portion of the SP field may be scattered out of the two-dimensional surface plasmon world into normal three-dimensional photons. These photons produced the image in Fig.4(b). We also conducted near-field optical measurements of the local surface plasmon field distribution around the droplet boundary Fig.4(c,d,e) using a sharp tapered optical fiber as a microscope tip. These measurements were performed similar to the measurements of surface plasmon scattering by individual surface defects described in \cite{18}. The droplet was illuminated with 488 nm laser light in the Kretschman geometry. The tip of the microscope was able to penetrate inside the glycerin droplet, and measure the local plasmon field distribution both inside and outside the droplet. Inside the droplet (in the right half of the images) the shear-force image (c) corresponds to the increase in viscous friction rather than the droplet topography. However, this image accurately represents the location of the droplet boundary, shown by the arrow in Fig.4(e). The sharp and narrow local maximum of the surface plasmon field just inside the droplet near its boundary is clearly visible in the near-field image Fig.4(d) and its cross-section Fig.4(e).         

Unfortunately, the described toy SP black hole model does not work outside the frequency range between $\omega _p/(1+\epsilon _d)^{1/2}$ and $\omega _p/2^{1/2}$. On the other hand, this is a common feature of every electromagnetic toy black hole model suggested so far. All such toy models necessarily work only within a limited frequency range. In addition, the losses in metal and dielectric described by the so far neglected imaginary parts of their dielectric functions will put a stop to the singularities of the field somewhere very near the toy event horizon (however, such "good" metals as gold and silver have very small imaginary parts of their dielectric constants, and hence, very pronounced plasmon resonances \cite{9}). Notwithstanding these limitations, the ease of making and observing such toy SP black holes makes them a very promising research object. If we forget about the language of "black holes" and "event horizons" for a moment, the SP optics phenomenon represented in Fig.4(b) remains a potentially very interesting effect in surface plasmon optics. Namely, this photo shows the existence of a two-dimensional SP analog of whispering gallery modes, which are well-known in the optics of light in droplets and other spherical dielectric particles. Whispering gallery modes in liquid microdroplets are known to substantially enhance nonlinear optical phenomena due to cavity quantum electrodynamic effects \cite{19}. One may expect even higher enhancement of nonlinear optical mixing in liquid droplets on the metal surfaces due to enhancement of surface electromagnetic field inherent to surface plasmon excitation, and in addition, due to accumulation of SP energy near the surface plasmon event horizons at the droplet boundaries. This strong enhancement of nonlinear optical effects in liquid droplets may be very useful in chemical and biological sensing applications. It may also be responsible for the missing orders of magnitude of field enhancement in the surface enhanced Raman scattering (SERS) effect \cite{12}, since various plasmon excitations are believed to play a major role in SERS. 

IV. NONLINEAR OPTICS OF SURFACE PLASMON BLACK HOLES.

The current explanation of SERS is based on the combination of electromagnetic and "chemical" enhancements \cite{12}. The current calculations of the electromagnetic field enhancement take into account "electrostatic" field enhancement at the apexes of various surface protrusions (the "lightning rod effect"), and the local field enhancement due to excitation of various localized surface plasmon modes in the crevices of the rough metal film. In addition, various weak and strong surface plasmon localization effects are considered in combination with the consideration of a rough metal surface as a fractal object \cite{20}. The "chemical" enhancement was proposed as an explanation for the quite a few missing orders of magnitude in theoretically calculated optical field enhancement, which follows from the magnitude of the experimentally measured SERS signals \cite{12}. The "chemical" enhancement may happen if the molecular energy levels are affected by the proximity to the rough metal surface, and are drawn into resonance with the excitation field. 

SERS observations are usually conducted under not so well controlled conditions when the surface topography is not well-known and "dirty" (the target molecules are present on the surface in random locations). On the other hand, it is well known that even monolayer surface coverages considerably shift plasmon resonance \cite{9} of the metal-vacuum interface. It is reasonable to suggest that some areas of the "dirty" metal surface may contain compact areas covered with the multiple layers of the target or solvent molecules (even if there are no droplets on the surface). Such compact areas would affect surface plasmon propagation in a way, which is very similar to the effect of the droplets considered above: effective surface plasmon event horizons may appear near the boundaries of such areas. Experimental data shown in Fig.4 and the theoretical arguments above strongly indicate that the local field enhancement near these boundaries may be considerable. Even the data shown in Fig.4(e) obtained with the limited optical resolution of the order of 100 nm \cite{18} indicate at least 10-fold enhancement of the square of the local field intensity near the droplet boundary. In the physical picture of surface plasmon whispering gallery modes the local SERS signal may be expected to grow by a factor of $Q^2$, where $Q$ is the cavity quality factor \cite{19}. For the surface plasmons in the visible range $Q$ may be estimated roughly as $Q\sim L/2\pi R$, where $L$ is the surface plasmon free propagation length and $R$ is the droplet radius. Taking into account the typical theoretical value of $L\sim 40\mu m$ \cite{21} in the visible range, $Q^2\sim 50$ may be obtained for a $R=1\mu m$ droplet. Thus, such surface plasmon "whispering galleries"/"black holes" may provide considerable SERS enhancements on top of the local electromagnetic enhancement due to other effects associated with the surface roughness.

The effects of surface roughness on the properties of plasmons near the effective event horizon have not been considered so far. However, detailed analysis of these effects conducted recently for the case of real black holes \cite{22} indicates that these effects will be especially strong in the nonlinear optics domain. For example, strong optical second harmonic generation has been predicted near the event horizon due to the effects of weak localization \cite{22}. 

Wave propagation and localization phenomena in random media have been the topic of extensive studies during the last years \cite{23}. One of the most striking examples of such phenomena is the strong and narrow peak of diffuse second harmonic light emission observed in the direction normal to a randomly rough metal surface (see Fig.5(a)). This peak is observed under the coherent illumination at any angle. This effect was initially predicted theoretically \cite{24} and later observed in the experiment \cite{25}. The enhanced second harmonic peak normal to the mean surface arises from the fact that a state of momentum {\bf k} introduced into a weakly localized system will encounter a significant amount of backscattering into states of momentum centered about {\bf -k}. When these surface {\bf k} and {\bf -k} modes of frequency $\omega $ interact through an optical nonlinearity to generate $2\omega $ radiative modes, the $2\omega $ light has nonzero wave vector components only perpendicular to the mean surface. The angular width of the normal peak can be as small as a few degrees, and its amplitude far exceeds the diffuse omnidirectional second harmonic background. 

Similar weak localization effects may be expected for the surface plasmon modes, which are trapped near the effective event horizon of our toy black hole when the metal surface exhibits moderate roughness. Similar to the case of planar rough surface \cite{24,25}, the momentum component parallel to the edge of the droplet must be conserved in nonlinear optical processes. This means that while the surface plasmon {\bf k} and {\bf -k} modes of frequency $\omega $, which interact through an optical nonlinearity, generate $2\omega $ modes, the $2\omega $ plasmons could have nonzero wave vector component only perpendicular to the mean edge of the droplet (Fig.5(b)). Thus, weak localization effects in the scattering of surface optical modes near the horizon should produce a pronounced peak in the angular distribution of second harmonics of plasmons in the direction normal to the horizon. In addition, because of such propagation direction (perpendicular to the edge of the horizon), these second harmonic plasmons have the best chances to escape the vicinity of the toy black hole. As a result, the relative intensity of the second harmonic radiation due to the weak localization effect near the event horizon will be much higher with respect to the diffuse SHG than in the case of planar rough surface. The diffuse SH plasmons would remain trapped near the horizon similar to the case of a real black hole \cite{22}.

Our experiments strongly indicate enhancement of SHG near the toy surface plasmon black holes. The droplet shown in Fig.6(a) was illuminated by the weakly focused beam (illuminated spot diameter on the order of 50 $\mu $m) from a Ti:sapphire laser system consisting of an oscillator and a regenerative amplifier operating at 810 nm (repetition rate up to 250 kHz, 100-fs pulse duration, and up to 10 $\mu $J pulse energy), which was directed onto the sample surface in the Kretschman geometry (Fig.1(a)). Excitation power at the sample surface was kept below the ablation threshold of the gold film. The local SHG from the droplet illuminated by the laser can be clearly seen in Fig.6(b), which was obtained using a far-field microscope and a CCD camera. The cross section (Fig.6(c)) of the second harmonic image indicates the droplet edge as an origin of SHG. It should be noted, however, that while the SH frequency plasmons experience the "event horizon" near the edge of the droplet, the refractive index of glycerine is not sufficiently large for the fundamental frequency plasmons to see it. Thus, 810 nm plasmons may only be trapped into the whispering gallery modes near the edge of the droplet. On the other hand, the basic weak localization mechanism remains the same in the studied experimental situation.

Finally, let us examine the potentially most interesting nonlinear optical phenomenon, which may be exhibited by our toy surface plasmon black holes. The Hawking radiation \cite{7} has been one of the most fascinating quantum physical phenomena discovered recently. However, nobody has observed it in the experiment so far, and its observation remains a great experimental challenge. In order to evaluate the chances to see it in the experiments performed with surface plasmon toy black holes, we must first derive the expression for their effective Hawking temperature. A derivation of an effective Hawking temperature of a toy dielectric black hole has been performed recently by Reznik \cite{1}. In the following we will use his results. According to the standard derivation procedure \cite{1,7,26}, we must introduce the analogue "Minkowski" coordinates as $U=-xe^{-\alpha t}$ and $V=xe^{\alpha t}$, which make the effective metric (4) conformally flat, and require that in the flat "Minkowski" space the plasmon field will be in its vacuum state $\mid 0_M>$. However, in the laboratory frame the "Minkowski" vacuum looks like a bath of thermal radiation with the temperature

\begin{equation}
T_H=\frac{M_{PL}^2c^2}{8\pi k_BM_{BH}}=\frac{\hbar c\alpha }{\pi k_B},
\end{equation}

where $M_{PL}=(\hbar c/\gamma )^{1/2}$ is the Planck mass and $k_B$ is the Boltzmann constant. For more details on the derivation of this result one may address \cite{1}, where it is also demonstrated that cutting off $c^\star $ near $x=0$ and inclusion of the dispersion does not eliminate the Hawking radiation. 

We should point out that the final result for the effective Hawking temperature does not depend on the gravitation constant $\gamma $. This is an encouraging fact since we are dealing with the toy black holes. In order to get a numerical estimate on the effective Hawking temperature we must assume some realistic value of $\alpha $. Our approximation of a slow adiabatically changing $c^\star $ inside the droplet means that $c^\star $ does not change considerably on the scale of the local wavelength $\lambda ^\star $ of surface plasmons. Thus, a good top estimate for $\alpha $ should look like $\alpha <1/\lambda \sim 1/\lambda _0$, where $\lambda $ is the plasmon wavelength far from the droplet, and $\lambda _0$ is the wavelength of light in vacuum. As a result, the effective Hawking temperature of our toy black hole is of the order of 

\begin{equation}
T_H\leq \frac{1}{2\pi ^2}\frac{\hbar \omega _p}{k_B}\sim 1000K
\end{equation} 

This value is quite close to the room temperature, which means that the droplet may probably conform itself to be exactly in thermal equilibrium with its ambient. Thus, we come to a surprising conclusion that the shape of liquid droplets on metal surfaces may to some degree be determined by the Hawking radiation. 

However, we should accept this conclusion with a degree of caution because of the recent result that the electromagnetic black hole analogues exhibit only the classical features of the black holes, while such quantum mechanical properties as Hawking radiation are not reproduced \cite{27}. The basic reason for not having thermal radiation emitted is the energy conservation: there should be a source of thermal energy emitted to infinity. On the other hand, we must remember that the surface plasmons are non-radiative modes, which are bound to the interface. In addition, they do not propagate far alone the interface, being absorbed by the metal within a few tens of micrometers from the source. As a result, the surface plasmon Hawking radiation associated with the toy black hole would be observable only in the near field of the black hole, and energy conservation would not be violated. This situation would be rather similar to the recently discovered near-field thermal surface phonon-polariton emission in SiC \cite{28}, when the monochromatic emission with photon energies not represented in the far field zone may be observed in the near field of the SiC surface. In fact, this interesting effect is caused by the fact that $c^\star \rightarrow 0$ under the conditions of surface phonon-polariton resonance, similar to the toy Hawking radiation effect considered above. In addition, one may create non-equilibrium situations when energy is pumped into a toy black hole. Under such non-equilibrium conditions Hawking radiation may also be observable in the experiment. 

V. SURFACE PLASMON WORMHOLES AND TIME MACHINES.

Looking back at the consequences of our model for gravitation theory, we may anticipate that one can look forward to modeling of more exotic gravitation theory situations where a wormhole exists inside a black hole. In the more prosaic language of surface plasmon optics this situation corresponds to nanoholes drilled in a free standing metal membrane and covered with droplets of dielectric (Fig.7(a,c)). In fact, our recent experiments on single-photon tunneling \cite{29} and optical control of photon tunneling \cite{30} were performed with exactly these kinds of samples. Theoretical analysis of these experiments performed in \cite{10} indicated some usefulness of the gravitational theory analogy: parallels can be drawn between the nonlinear optics of surface plasmons in cylindrical nanoholes (cylindrical surface plasmons) and areas of gravitation theory such as Kaluza-Klein theories \cite{10}. 
  
This analogy stems from the way in which electric charges are introduced in the original five-dimensional Kaluza-Klein theory (see for example \cite{31,32} and Fig.7(b)). In this theory the electric charges are introduced as chiral (nonzero angular momentum) modes of a massless quantum field, which is quantized over the cyclic compactified fifth dimension. Electromagnetic forces between the electric charges appear as nonlinear coupling of these chiral modes, so that four-dimensional electrodynamics described by the Maxwell equations may be understood as nonlinear optics of these modes. Similar Kaluza-Klein theories may be formulated in lower-dimensional space-times. Such theories reproduce electrodynamics of electric charges in worlds with less than three spatial dimensions. Since nonlinear interactions of surface plasmons of a cylindrical nanowire or a nanohole look as if they occur in a space-time which besides an extended z-coordinate has a small "compactified" angular $\phi $-dimension along the circumference of the cylinder, the theory of cylindrical surface plasmon (CSP) mode propagation and interaction appears to be similar to the three-dimensional Kaluza-Klein theory: solutions of the nonlinear Maxwell equations for interacting cylindrical surface plasmons may be found \cite{10}, which behave as interacting Kaluza-Klein charges. According to these solutions, higher $(n>0)$ CSP modes posses quantized effective chiral charges proportional to their angular momenta $n$. In a metal nanowire these slow moving effective charges exhibit long-range interaction via exchange of fast massless CSPs with zero angular momentum. These zero angular momentum CSPs may be considered as massless quanta of the gyration field (the field of the gyration vector $\vec{g}$, see eq.(9)), which relates the $\vec{D}$ and $\vec{E}$ fields in an optically active medium \cite{15}. This is possible because of the magneto-optical effect (eq.(10)) and the fact that the magnetic field of the massless $n=0$ CSP modes points in the angular direction.

In a thin metal wire such interaction of the CSP chiral charges via exchange of zero-angular-momentum CSPs becomes quite noticeable at large distances. Unlike the Coulomb interaction of electrons, which is screened by the presence of other free electrons, the chiral interaction does not experience much screening at short distances, due to its general weakness: we may say that the chiral charges are adiabatically free, similar to the behavior of quarks at short distances. However, because of the one-dimensional nature of the chiral interaction in a long nanowire, the force of interaction does not depend on the distance between the chiral charges unless the CSP decay (finite free propagation length due to losses in metal) is taken into account. This fact may make the chiral interaction an important mechanism in mesoscopic transport phenomena, especially in long samples.

In order for such a physical picture to be valid the medium in or around the nanohole or nanowire should be chiral or optically active, and exhibit the magneto-optical effect. This condition is fulfilled automatically in the case of any metal \cite{15}, since all metals are optically active in the presence of a magnetic field. Thus, optical experiments performed on nanowires and nanoholes may become the proving ground for experimental testing of many ideas in theoretical physics, such as compactified extra dimensions and wormholes.

Detailed illustration of how the nonlinear optics of cylindrical surface plasmons may be formulated in a way that is similar to Kaluza-Klein theories can be found in \cite{10}. The effective metric of a surface plasmon wormhole can be derived as follows. Let us represent a thin cylindrical metal wire (or hole) by considering a four-dimensional space-time, which besides the extended z-coordinate has two compactified cyclic spatial dimensions: the $\phi $-dimension along the circumference of the metal cylindrical wire (hole), and the cyclic fifth $\theta $-dimension of the original Kaluza-Klein theory. The effects of the radial coordinate will be neglected for the sake of simplicity (some justification of it can be found in \cite{10}). We will also assume temperature to be very low, so that interactions of only a few electrically charged quaziparticles can be considered, while the rest of the electrons in the metal are taken into account via electric and magnetic permeability tensors $\epsilon _{ik}$ and $\mu _{ik}$ of the metal, which in turn are taken into account via the effective space-time metric $g_{ik}$. Thus, the effective metric can be written as

\begin{eqnarray}
ds^2 = c^2dt^2 - dz^2 - R^2d\phi ^2 - r^2d\theta ^2+ 2g_{02}cdtd\phi + 
\nonumber \\
2g_{03}cdtd\theta + 2g_{12}dzd\phi + 2g_{13}dzd\theta + 2g_{23}d\theta d\phi
\end{eqnarray} 

where $R(z)$ and $r$ are the radii of the cylindrical wire and the compactified original fifth Kaluza-Klein dimension, respectively. Here we consider all the $g_{0i}$ and $g_{1i}$ components to be independent of $\phi $ and $\theta $. An infinite nanohole or nanowire geometry would be described by $R(z)=const$, while a toy wormhole geometry in Fig.7(a) corresponds to $R(z)=\infty $ at $z=\pm l/2$, where $l$ is the wormhole length. The field equations for a quantum massless scalar field $\psi $ in this metric should be written as 

\begin{equation}
\frac{\partial }{\partial x^i}(g^{ik}\frac{\partial \psi }{\partial x^k})=0,
\end{equation}

where $i,k=0, ... , 3$. The field $\psi $ is considered to be scalar in order to make the consideration as simple as possible. If we search for the solutions in the usual form as
$\psi = \Psi (x^{\alpha })e^{iq_2\phi }e^{iq_3\theta }$, where $\alpha =0,1$, and assume periodicity in $\phi $ and $\theta $, we obtain $q_2=n_2$ and $q_3=n_3$ (where $n_2$ and $n_3$ are integer). As a result, equation (15) takes the form:

\begin{eqnarray}
(\frac{\partial ^2}{c^2\partial t^2}-\frac{\partial ^2}{\partial z^2})\psi -(q_2^2g^{22}+q_3^2g^{33}+2q_2q_3g^{23})\psi+ \nonumber \\ 
2iq_2g^{\alpha 2}\frac{\partial \psi}{\partial x^{\alpha }}+2iq_3g^{\alpha 3}\frac{\partial \psi}{\partial x^{\alpha }}+iq_2 (\frac{\partial g^{\alpha 2}}{\partial x^{\alpha }})\psi+iq_3 (\frac{\partial g^{\alpha 3}}{\partial x^{\alpha }})\psi = 0
\end{eqnarray} 

This is a two-dimensional Klein-Gordon equation describing a two-component quantized charge $(q_2,q_3)$ in the presence of external $g^{\alpha 2}(t,z)$ and $g^{\alpha 3}(t,z)$ vector fields. The $q_3$ component of the charge and the $g^{\alpha 3}$ field correspond to the quantized electric charge $e \sim q_3$ and the electromagnetic field, respectively, while the $q_2$ component of the charge and the $g^{\alpha 2}$ field correspond to the angular momentum of the plasmon (the chiral charge) and the gyration field described above (eq. (9)). 

The value of this Kaluza-Klein wormhole and the black hole models for the field of nanooptics is quite apparent: these models provide very useful intuitive guides for finding solutions of nonlinear Maxwell equations in the situations when nonlinear interactions are strong and can not be considered as small perturbations. In such situations, often one can only guess the general form of the solutions, based on the comparison of the nonlinear system of interest with other better understood nonlinear systems described by similar equations. We can also ask if experimental and theoretical studies of nanoholes and microdroplets can be useful for gravitation theory. Some arguments can be made that this might be the case. For example, the solutions for the electromagnetic eigenmodes of the wormholes and their interaction are often sought for very narrow wormholes, which are understood as strings connected to D-branes (D-dimensional membranes) \cite{33}. The multi-component Kaluza-Klein charges of the wormholes in such configurations have some resemblance to the multi-component chiral charges of metal nanowires and nanoholes described above. Thus, solid state models may provide some useful insights. It is customary in theoretical description of wormholes in higher-dimensional Kaluza-Klein theories to distinguish between electric and magnetic Kaluza-Klein charges and fields \cite{34}. According to this classification, we should identify the chiral charges of the cylindrical surface plasmons as magnetic charges, while (quite naturally) the conductive electrons in the metal nanowire perform the role of the electric Kaluza-Klein charges \cite{10}. Another important question, which follows from the analogy between the nanoholes and the wormholes arises from the fact that many optical waveguides (from the electrodynamics point of view a wormhole is a waveguide) have cut-off frequencies for the propagating electromagnetic modes. For example, unlike metal nanowires, the nanoholes do not support propagating electromagnetic modes with small frequencies. This may mean that in gravitation theory an attempt to describe a wormhole using a frequency-independent space-time metric may fail in some cases. Excitations with small energies below the wormhole cut-off may not see the wormhole throat at all.      

In addition, consideration of toy black holes and wormholes may allow us to take another look at such a longstanding and controversial gravitation theory subject as chronology protection conjecture. Real wormholes and rotating black holes are the basic elements which may allow, in principle, a creation of a time machine according to a number of published designs \cite{35,36}. Let us try and emulate these time machine designs with toy surface plasmon "black holes" and "wormholes". Assuming that such a toy "time machine" does not work, a general prediction can be made on strong electromagnetic field enhancement inside an arbitrary-shaped nanohole near an arising effective event horizon, which is supposed to prevent the surface plasmon "time machine" from being operational. This general result is useful in description of the nonlinear optical behavior of random and artificial nanoholes in metal films, which indeed show signs of strongly nonlinear behavior in recently observed single-photon tunneling and optically-controlled photon tunneling experiments \cite{29,30}.

Let us recall the basic principles of a "real" time machine operation. In a time machine design of Morris, Thorne and Yurtsever \cite{35} (Fig.8(a)) at the initial time the openings of the wormhole are not far apart. Assuming that one of the openings is forced to move away at high speed (while internal distance between the openings remains unchanged and short at all times), and then to reverse its motion and return back to the vicinity of the second opening, upon completion of the motion the clock of the moving opening will lag the clock of the stationary wormhole opening. As a result, an observer passing through the wormhole may travel to his past or future depending on the direction of travel through the wormhole. Another variant of the time machine introduced by Novikov \cite{36} (Fig.8(b)) consists of a wormhole in which one opening rotates around another. The clock of a moving opening lags the clock of a stationary one, so that after some time a Cauchy horizon appears, and travel into the past becomes possible. 

Now let us discuss the properties of the local "proper" time $\tau = c^\star t$, which appears in the effective metric of surface plasmon black holes (eq.(3,4)). This "proper" plasmon time at a metal-dielectric interface (for example, deep inside a liquid droplet) always lags the "proper" time at a metal-vacuum interface, due to the slower surface plasmon phase velocity. As a result, addition of a dielectric near metal surface emulates the slowing down of the clocks due to the motion of a reference frame or due to the gravitation field. Thus, we can try and emulate the motion of the wormhole openings in one of the time machine designs shown in Fig.8 by placing a droplet of dielectric on top of our toy wormhole (which means placing a liquid droplet (Fig.9a) on top of a hole drilled in a metal membrane (Fig.7a,c)). A plasmon trajectory starting from the top side of the membrane near the droplet edge and bringing it to its bottom side, and back to the top through the hole would emulate a time traveler motion described in \cite{35,36}. However, even though we would be using a distorted model "proper" time $\tau = c^\star t$ in this experiment, and there is no actual slowing down of a real clock of an observer performing experiments with surface plasmons, we are not supposed to build a surface plasmon toy "time machine". The surface plasmons cannot go back in time even according to the readings of a clock measuring this distorted "proper" time. This is clear from the fact that $c^\star $ is always positive. 

After a careful consideration we notice though, that the nature prevents operation of such a time machine: inside the hole drilled in the metal membrane another effective horizon is bound to appear. In the case of a smooth cylindrical hole partially filled with the dielectric (Fig.9b), this is quite clear from the consideration of dispersion of cylindrical surface plasmons \cite{37}. At large wavevectors ($k>>1/R$, where R is the cylinder radius) the dispersion law of cylindrical plasmons looks similar to the dispersion of regular surface plasmons (eq.(1)). Thus, virtually the same consideration as above may be repeated for cylindrical plasmons near the meniscus boundary, and we come to a conclusion that an effective horizon should appear inside the meniscus. 

However, the impossibility of building a time machine implies a much stronger conclusion, which would be valid in any arbitrarily-shaped pinhole in a metal film covered with a dielectric: an effective horizon must appear in any such pinhole. The value of this conclusion is in its generality. Maxwell equations are not easy to solve for some random, asymmetric, arbitrarily-shaped pinholes. The consideration above predicts strong electromagnetic field enhancement in any such geometry, which means that nonlinear optical properties of pinholes covered with dielectrics should be strongly enhanced. This conclusion agrees well with our experimental observations \cite{29,30}.

Nonlinear optical effects put limits on the validity of the effective metrics (3) or (4). While, this statement is true in general, it may not be valid for the lowest order nonlinearities. For example, the nonlinearities of the liquid dielectric of the form

\begin{equation}
\epsilon _d=\epsilon ^{(1)}_d+4\pi \chi ^{(3)}E^2 ,
\end{equation}

where $\epsilon ^{(1)}_d$ is the linear dielectric constant, $\chi ^{(3)}$ is the third-order nonlinear susceptibility of the liquid, $E$ is the local electric field, and $\chi ^{(3)}>0$, which would be responsible for the self-focusing effect in three-dimensional optics, may lead to a toy "gravitational collapse" of the surface plasmon field near the toy "black hole". This type of nonlinearity causes an effective gravitational interaction of surface plasmons with each other (here we consider the case of a central-symmetric nonlinear liquid with $\chi ^{(2)}=0$). Thus, we may imagine a situation where a liquid droplet is illuminated with an intense plasmon beam at a frequency below $\omega _p/(1+\epsilon ^{(1)}_d)^{1/2}$, so that a low intensity plasmon field would not experience an event horizon near the droplet edge. However, the increase in the droplet refractive index due to the high intensity plasmon field will cause the plasmon field to collapse towards an arising event horizon. Such a self-focusing effect may cause even stronger local field enhancement in the droplet and nanohole geometries observed in our experiments \cite{29,30}. 

In conclusion, we have introduced and observed experimentally surface plasmon analogues of such nontrivial space-time topologies as black holes and wormholes. Toy surface plasmon black holes are shown to exhibit strongly enhanced nonlinear optical behavior in the frequency range near the surface plasmon resonance of a metal-liquid interface. This enhancement may be responsible for the missing orders of magnitude of field enhancement in the surface enhanced Raman scattering effect. In addition, experimental observation of the recently predicted optical second harmonic generation near the toy event horizon has been reported. Finally, the possibility of toy Hawking radiation observation has been discussed, and the expression for the effective Hawking temperature of a toy surface plasmon black hole has been derived. We anticipate that further consideration of such analogues will produce a mutually beneficial exchange of ideas between nanooptics and gravitation theory. 

This work has been supported in part by the NSF grants ECS-0210438 and ECS-0304046.

Figure captions.

Fig.1 (a) Experimental geometry of surface plasmon toy black hole observation. (b) Surface plasmon dispersion laws for the cases of metal-vacuum interface far from the droplet, for metal-dielectric interface deep inside the droplet, and somewhere near the droplet edge. (c) A surface plasmon trapped inside a droplet near the effective "event horizon": The projection of surface plasmon momentum parallel to the droplet edge must be conserved. Due to effectively infinite refractive index near the droplet edge surface plasmons experience total internal reflection at any angle of incidence. 

Fig.2 (a) A thin metal membrane with two "linear" droplets positioned symmetrically on both sides of the membrane. An effective event horizon is located at $x=0$. Such geometry emulates the effective space-time geometry considered in \cite{1}. (b) Dispersion laws of the $\omega _-$ and $\omega _+$ modes exhibit positive and negative dispersion, respectively, near the surface plasmon resonance. These branches are shown for the cases of metal-vacuum interface far from the droplets, and for the locations near $x=0$.  

Fig.3 (a) Surface plasmon dispersion law for the metal-vacuum and metal-dielectric interfaces in the case of optically active dielectric. (b) Difference in refractive indices for the left and right surface plasmons results in different locations of the effective event horizons for these plasmons. The local directions of the gyration filed are shown by the arrows.

Fig.4 Far-field (a,b) and near-field (c,d) images of a toy surface plasmon black hole: (a) Droplet of glycerin on a gold film surface (illuminated from the top). The droplet diameter is approximately 15 micrometers. (b) The same droplet illuminated with white light in the Kretschman geometry, which provides efficient coupling of light to surface plasmons on the gold-vacuum interface (Fig.1(a)). The white rim around the droplet boundary corresponds to the effective surface plasmon "event horizon". (c) and (d) show $10\times 10 \mu m^2$ topographical and near-field optical images of a similar droplet boundary (droplet is located in the right half of the images) illuminated with 488 nm laser light. Cross-sections of both images are shown in (e). Position of a droplet boundary is indicated by the arrow.  

Fig.5 (a) Strong enhancement of second harmonic emission from a randomly rough metal surface in the direction normal to the surface. A typical angular distribution of the diffuse second harmonic light is shown in the inset. (b) Second harmonic generation near the toy black hole horizon. Surface roughness is represented by a rough droplet edge. Second harmonic plasmons emitted perpendicular to the horizon have the best chances to escape the toy black hole.

Fig.6 (a) Microscopic image of the glycerin droplet under normal illumination. (b) SHG is seen from the droplet illuminated by 810 nm Ti:sapphire laser light in the Kretschman geometry shown in Fig.1(a). The cross section (c) of the second harmonic image (b) indicates the droplet edge as an origin of SHG.

Fig.7 (a) A nanohole in a metal membrane may be treated as a wormhole for surface plasmons which exist on the flat top and bottom interfaces of the membrane. Alternatively, this figure may represent a nanowire connecting two metal-vacuum interfaces. (b) Similarity between Kaluza-Klein electric charges and cylindrical surface plasmons. (c) Array of 20 nm diameter nanoholes drilled in a free standing gold membrane, which has been studied experimentally. Such nanoholes may be treated as surface plasmon wormholes.

Fig.8 (a) Time machine design of Morris, Thorne and Yurtsever. Wormhole opening A is stationary, while the opening B moves along the Z-direction and returns back. (b) Time machine design of Novikov. Wormhole opening A is stationary, while the opening B rotates around it. 

Fig.9 (a) Surface plasmon toy "time machine": a droplet of dielectric is placed on top of the hole drilled in a metal membrane near its opening B, in order to emulate its motion. The metal membrane is made thicker to the right of the hole in order to allow plasmon propagation from the top side of the membrane to its bottom side, and back to the top through the hole (opening A). (b) Cylindrical plasmons inside a smooth cylindrical channel partially filled with dielectric experience an effective event horizon at the edge of the meniscus.

\end{document}